\documentclass[pra,aps,twocolumn,superscriptaddress,showpacs]{revtex4}
\usepackage{epsfig,amsmath}

\begin{document}
\title{Continuous Multipartite
Entangled State in Wigner Representation and the Violation of
\.{Z}ukowski-Brukner Inequality}

\author{Chunfeng Wu}
\email{g0201819@nus.edu.sg}
\affiliation{Department of Physics, National University of
Singapore, 2 Science Drive 3, Singapore 117542}
\author{Jing-Ling Chen}
\email{phycj@nus.edu.sg}
\affiliation{Department of Physics, National University of
Singapore, 2 Science Drive 3, Singapore 117542}
\author{L. C. Kwek} 
\affiliation{Department of Physics, National University of
Singapore, 2 Science Drive 3, Singapore 117542}
\affiliation{National Institute of Education, Nanyang
Technological University, 1 Nanyang Walk, Singapore 639798 }
\author{C. H. Oh}
\affiliation{Department of Physics, National University of
Singapore, 2 Science Drive 3, Singapore 117542}
\author{Kang Xue}
\affiliation{ Department of Physics, Northeast Normal University,
Changchun 130024, P. R. China }
\newcommand{\ssb}{{\cal B}}

\pacs{03.65.-w,42.50.Dv}

\begin{abstract}
We construct an explicit Wigner function for $N$-mode squeezed
state. Based on a previous observation that the Wigner function
describes correlations in the joint measurement of the phase-space
displaced parity operator, we investigate the non-locality of
multipartite entangled state by the violation of
\.{Z}ukowski-Brukner $N$-qubit Bell inequality. We find that
quantum predictions for such squeezed state violate these
inequalities by an amount that grows with the number $N$.
\end{abstract}
\maketitle

Einstein, Podolsky and Rosen (EPR)challenged the completeness of
quantum mechanics in a classic seminal paper in 1935\cite{EPR}.
In this paper, they reasoned that the wavefunction of a
two-particle system in which the particles are entangled in
position and momentum (an EPR pair) and written explicitly as
\begin{eqnarray}
\Psi(q_1, q_2)=\int_{-\infty}^{\infty}e^{(2\pi
i/h)(q_1-q_2+q_0)p}dp \label{EPR1}
\end{eqnarray}
is incompatible with the completeness postulate. It was
subsequently argued that additional variables (local hidden
variables) could be introduced to restore casuality and locality
to quantum mechanics.

A scheme \cite{Bell1} for testing the compatibility of the theory
of local hidden variables with quantum mechanics was subsequently
proposed based on a different version of an EPR pair based on the
entanglement of spin-$1/2$ particles first introduced by Bohm. In
particular, the possibility of local realism implies logical
constraints on the statistics of two or more physically separated
systems. These constraints can be expressed in the form of
Bell-type inequalities \cite{Bell1, CH, CHSH, Green, Peres, JLC,
Zukowski}. For quantum mechanical systems, it was anticipated
that these constraints could be violated with appropriate
measurements.

On the other hand, quantum correlations for position-momentum
variables can be analyzed in position-momentum phase space using
the Wigner function \cite{Wigner}. The Wigner function allows one
to define a probability distribution in position-momentum phase
space for a quantum mechanical particle. This formalism led
eventually to the formulation of a c-number approach to describe
the quantum effects in phase space including the development of
various other efficient tools in a number of fields in modern
physics \cite{Wigner1}.

Indeed, a mixed state system can be represented by a density
matrix ${\bf{\rho}}$
\begin{eqnarray}
P_W(q,p)=\frac{1}{\pi\hbar}\int_{-\infty}^{\infty}dy\langle
q-y|{\bf{\rho}}|q+y\rangle e^{\frac{2ipy}{\hbar}}
\end{eqnarray}
where ${\bf{\rho}}$ is a density operator, $|q\rangle$ is the
eigenvector of the coordinate operator. An intuitive physical
meaning of the Wigner function is that its marginal distribution
in one of the variables gives the probability distibution of the
particle in that space. For momentum, we have
$\displaystyle P_{mom}(p)=\int W(q,p)dq $ and with position $
\displaystyle P_{pos}(q)=\int W(q,p)dp $.

Bell\cite{Bell2} had argued that the original EPR wave function
does not violate local realism because its joint Wigner distribution
function $W(q_1,p_1;q_2,p_2)$ is positive everywhere, and as such
it will also admit a local hidden variable description of signed
position correlations. However, the  choice of  appropriate
observables is important for  testing the existence of local
realism for a given state. In a recent work, Banaszek and
W\'{o}dkiewicz \cite{KK} considered parity measurement, a quantum
observable which does not admit a local hidden variable
description,  and interpreted the Wigner function as a
correlation function for these parity measurements. They then
showed that the original EPR state and the two-mode squeezed
vacuum state violate local realism since they violate generalized
Bell inequalities such as the Clauser-Horne inequality \cite{CH}
and the Clauser-Horne-Shimony-Holt (CHSH)\cite{CHSH} inequality.
In particular, they considered two-mode squeezed vacuum state
produced through non-degenerate optical parametric amplification
(NOPA)\cite{NOPA} in order to avoid problems related to the
singularity of the unnormalizable EPR state. Moreover, in Ref.
\cite{KK}, it was shown that despite its positive definiteness,
the Wigner function of the EPR state could provide direct
evidence of the non-locality.

The two-mode squeezed vacuum state generated in a nondegenerate
optical parametric amplifier (NOPA)\cite{NOPA} is given by
\begin{equation}
| {\rm NOPA}\rangle =e^{r(\hat{a}_1^{\dagger }\hat{a}_2^{\dagger
}-\hat{a}_1\hat{a}_2)}|00\rangle =\sum_{n=0}^\infty \frac{(\tanh r)^n}{\cosh r}%
|nn\rangle ,  \label{nopa}
\end{equation}
where $r$ is known as the squeezing parameter and $\left|
nn\right\rangle
\equiv \left| n\right\rangle _1\otimes \left| n\right\rangle _2=\frac 1{n!}%
(\hat{a}_1^{\dagger })^n(\hat{a}_2^{\dagger })^n\left|
00\right\rangle $. The NOPA states $\left| {\rm
NOPA}\right\rangle $ can also be written as \cite{KK}:
\begin{equation}
\left| {\rm NOPA}\right\rangle =\sqrt{1-\tanh ^2r}\int dq\int
dq^{\prime }g(q,q^{\prime };\tanh r)|qq^{\prime }\rangle ,
\label{epr}
\end{equation}
where $g\left( q,q^{\prime };x\right) \equiv \frac 1{\sqrt{\pi
(1-x^2)}}\exp
\left[ -\frac{q^2+q^{\prime 2}-2qq^{\prime }x}{2(1-x^2)}\right] $ and $%
\left| qq^{\prime }\right\rangle \equiv \left| q\right\rangle
_1\otimes \left| q^{\prime }\right\rangle _2$, with $\left|
q\right\rangle $ being the eigenstates of the position operator.
Since $\lim_{x\rightarrow
1}g\left( q,q^{\prime };x\right) =\delta (q-q^{\prime })$, one has $%
\lim_{r\rightarrow \infty }\int dq\int dq^{\prime }g(q,q^{\prime
};\tanh r)|qq^{\prime }\rangle =\int dq|qq\rangle =\left| {\rm
EPR}\right\rangle $, which is just the original EPR states. Thus,
in the infinite squeezing limit, $\left| {\rm NOPA}\right\rangle
\left| _{r\rightarrow \infty }\right. $ becomes the original,
normalized EPR states. Since the original EPR state is an
unnormalized $\delta$ function, a normalizable state generated in
a NOPA can avoid problem arising from this singularity.

The Wigner function can be associated directly with the parity
operator. The connection between the parity operator
$(-1)^{\hat{n}}$ and the Wigner function provides an equivalent
definition of the latter\cite{BGE1}. The Wigner representation of
the parity operator is not a bounded reality corresponding to the
dichotomic result of the measurement. This enables violation of
Bell inequality for quantum states described by positive-definite
Wigner function.

The relation between the non-locality of arbitrary multipartite
entangled states and the Wigner function remains an open
question. Recently, tripartite entangled state representation of
the Wigner operator and the corresponding Wigner function have
been found by Fan and Jiang \cite{FAN}. They focused principally
on a generalization of the Wigner function and its marginal
distributions, without invoking the non-locality issue. A general
Bell inequality which is a sufficient and necessary condition for
the correlation function for $N$ particles has been described in
Ref. \cite{Zukowski}. In this work, measurements on each particle
were chosen from two arbitrary dichotomic observables. Namely, the Zukowski-Brukner inequality was derived for observables with eigenvalues $+1$ or $-1$, which are also the spectrum of the displaced parity operator.
Thus this general Bell theorem for general N-qubit states provides a useful
tool to test the violation of local realism of multipartite
quantum states described by Wigner function. With this
motivation, we derive an expression for the Wigner function of
N-mode squeezed state in this Letter. By expressing the
correlation function using the Wigner function, we show that the
multipartite entangled state violates local realism, and this
violation is enhanced with increasing dimension, $N$.

To this end, we first choose parity operators as the observables
for testing violation of local realism for a squeezed state. The
Wigner function can be expressed as the expectation value of a
product of displaced parity operators as follows
\begin{eqnarray}
W(\alpha_1, \alpha_2,..., \alpha_N)\propto\Pi(\alpha_1,
\alpha_2,..., \alpha_N)
\end{eqnarray}
where $\Pi(\alpha_1,\alpha_2,..., \alpha_N)$ is the expectation value of the joint displaced parity operator (i.e. the measured observable) 
\begin{eqnarray}
 \hat{\Pi}(\alpha_1, \alpha_2,..., \alpha_N) &= & \hat{D}_1(\alpha_1)...\nonumber \\
 & & \mbox{\hspace{-3cm }} \times
 \hat{D}_N(\alpha_N)(-1)^{\hat{n}_1+...
 +\hat{n}_N}\hat{D}^{-1}_N(\alpha_N)...\hat{D}^{-1}_1(\alpha_1)\label{displaced}
\end{eqnarray}

In the above expression,
$\hat{D}_i(\alpha_i)=\exp(\alpha_i\hat{a}^{\dagger}-\alpha_i^*\hat{a})$
denotes the displacement operators for the subsystem $i$, where
$\hat{a}(\hat{a}^\dagger)$ is annihilation (creation) operator.
We also equate the correlation functions given by the displaced
parity operator $(-1)^{\hat{n}_1+...+\hat{n}_N}$ as the
equivalent Wigner function \cite{parity}. In this way, we see
that the non-local realsitic description is embedded in the
dichotomic correlation measurements given by the phase-space
Wigner function for the entangled state,
\begin{eqnarray}
E(\alpha_1, \alpha_2,..., \alpha_N)\equiv\Pi(\alpha_1,
\alpha_2,..., \alpha_N)
\end{eqnarray}

The Wigner function, or equivalently the correlation function for
multipartite system, can be calculated using the expectation value
of the operator under the $N$-mode squeezed state. This new
squeezed state, a $SU(1,1)$ coherent state, is given as
\begin{eqnarray}
|r\rangle=V|\textbf{0}\rangle=\exp[r(W_+-W_-)]|\textbf{0}\rangle
\end{eqnarray}
where $|\textbf{0}\rangle=|00...0\rangle$ is a N-mode vacuum
state, and
\begin{eqnarray}
W_+&=&x\sum^{N}_{i=1}\hat{a}_i^{\dagger 2}+y\sum^{N}_{i<j=1}\hat{a}_i^\dagger \hat{a}_j^\dagger \nonumber \\
W_-&=&x\sum^{N}_{i=1}\hat{a}_i^2+y\sum^{N}_{i<j=1}\hat{a}_i\hat{a}_j \nonumber \\
B&=&\frac{1}{2}\sum^{N}_{i=1}\hat{a}^\dagger _i\hat{a}_i+\frac{N}{4}
\end{eqnarray}
$W_+$ is a $N$-mode squeezing operator and $x$ and $y$ are the
coefficients which can be determined by the fact that the above
formula satisfies the closed $SU(1,1)$ Lie algebra:
$ [W_+,W_-]=-2B, [W_+,B]=-W_+, [W_-,B]=W_-$. The final result is
\begin{eqnarray}
W_+=\frac{2-N}{2N}\sum^{N}_{i=1}\hat{a}_i^{\dagger 2}+\frac{2}{N}\sum^{N}_{i<j=1}\hat{a}_i^\dagger \hat{a}_j^\dagger \\
W_-=\frac{2-N}{2N}\sum^{N}_{i=1}\hat{a}_i^2+\frac{2}{N}\sum^{N}_{i<j=1}\hat{a}_i\hat{a}_j
\end{eqnarray}

The $N$-mode squeezed state is characterized by the squeezing
parameter $r$. The Wigner function of the squeezed state is
calculated in the following way. When $r$ is zero, namely when no
squeezing occurs, the Wigner function is given by
\begin{eqnarray}
E(\alpha_1, \alpha_2,..., \alpha_N)&=&\langle \textbf{0}
|\hat{\Pi}(\alpha_1, \alpha_2,..., \alpha_N)|\textbf{0}\rangle \nonumber \\
&=&\exp[-2\sum^{N}_{i=1}|\alpha_i|^2]
\end{eqnarray}
When $r\not=0$, the new Wigner function can be constructed from
\begin{eqnarray}
E'(\alpha_1, \alpha_2,..., \alpha_N)&=&
\langle r|\hat{\Pi}(\alpha_1, \alpha_2,..., \alpha_N)
|r\rangle \nonumber \\
&=&\langle \textbf{0}|\hat{\Pi}(\alpha'_1, \alpha'_2,..., \alpha'_N)|\textbf{0}\rangle \nonumber \\
&=&\exp[-2\sum^{N}_{i=1}|\alpha'_i|^2]
\end{eqnarray}
where
$\hat{\Pi}(\alpha'_1, \alpha'_2,..., \alpha'_N)$is the squeezed
displaced parity operator given in Eq.(\ref{displaced}). After some
lengthy calculation, we arrive at the following relations,
\begin{eqnarray}
V^{-1}\hat{a}_iV=\cosh r\hat{a}_i+\sinh r(\frac{2-N}{N}\hat{a}_i^\dagger+\frac{2}{N}\sum^{N}_{j\not=i}\hat{a}_j^\dagger) \\
V^{-1}\hat{a}^\dagger_iV=\cosh r\hat{a}^\dagger_i+\sinh
r(\frac{2-N}{N}\hat{a}_i+\frac{2}{N}\sum^{N}_{j\not=i}\hat{a}_j).
\end{eqnarray}
The latter relation can be employed to yield the Wigner function
of N-mode squeezing state
\begin{eqnarray}
E'(\alpha_1, \alpha_2,..., \alpha_N)&=& \exp\{-2\cosh 2r\sum_{i=1}^{N}|\alpha_i|^2 \nonumber \\
& & \mbox{\hspace{-2cm}}+\frac{4}{N}\sinh2r\sum^{N}_{i<j}(\alpha_i\alpha_j+\alpha_i^*\alpha_j^*)\nonumber \\
&&\mbox{\hspace{-2cm}}
-\frac{N-2}{N}\sinh2r\sum^{N}_{i=1}(\alpha_i^2+\alpha_i^{*2})\}
\end{eqnarray}
The Wigner function of the original EPR state is recovered in the
limit of $r\rightarrow\infty$ for $N=2$.

The $N$-NOPA field modes are equivalent to an entangled state of
$N$ oscillators. When $N=3$, the Wigner function is
\begin{eqnarray}
E'(\alpha_1, \alpha_2,\alpha_3)&=&\exp\{-2 \cosh2r\sum^{3}_{i=1}
|\alpha_i|^2\nonumber \\
& & \mbox{\hspace{-2cm}}+\frac{4}{3}\sinh 2r\sum^{3}_{i<j}(\alpha_i \alpha_j+\alpha_i^*\alpha_j^*)\nonumber \\
&& \mbox{\hspace{-2cm}}
-\frac{1}{3}\sinh2r\sum^{3}_{i=1}(\alpha_i^2+\alpha_i^{*2})\}
\end{eqnarray}
and this is the same as the result given in Ref. \cite{FAN}. The
correlation function is determined by considering measurements
corresponding to the settings $\alpha_1^1=0,\alpha_1^2=a$,
$\alpha_2^1=0,\alpha_2^2=a$, and $\alpha_3^1=-a,\alpha_3^2=0$, where $\alpha_i^j$ (i=1,2,3 and j=1,2) is the jth measurement setting for ith particle, and $a$ is a
positive constant associated with the displacement magnitude.
From these combinations, the following quantity can be constructed
\begin{widetext}
\begin{eqnarray}
\ssb(3)&&=E'(\alpha_1^1,\alpha_2^1,\alpha_3^2)+E'(\alpha_1^1,\alpha_2^2,\alpha_3^1)+E'(\alpha_1^2,\alpha_2^1,\alpha_3^1)-E'(\alpha_1^2,\alpha_2^2,\alpha_3^2)\nonumber \\&&
=E'(0,0,0)+E'(0,a,-a)+E'(a,0,-a)-E'(a,a,0)\nonumber \\&&
=1+2\exp\{(-4\cosh2r-\frac{8}{3}\sinh2r-\frac{4}{3}\sinh2r)a^2\}-\exp\{(-4\cosh2r+\frac{8}{3}\sinh2r-\frac{4}{3}\sinh2r)a^2\}
\label{eqn1}
\end{eqnarray}
\end{widetext}
For local hidden variables theories, we have the inequality
\cite{Zukowski} $-2\leq\ssb(3)\leq2$. If we perform an asymptotic
analysis for large $|r|$ with $r<0$, $\cosh2r$ and $\sinh2r$ can be replaced by
$e^{-2r}/2$ and $-e^{-2r}/2$ respectively, and Eq.(\ref{eqn1})
becomes $ \ssb(3)=3-\exp\{-\frac{8}{3}e^{-2r}a^2\} $ We see that
when $a^2/e^{2r}$ is large enough, the Bell inequality for three
qubits is violated when $\ssb(3)$ approaches the value
$\ssb_{\mbox{opt}}=3$.

For $N=4$, and choosing all $\alpha_i$ to be reals, the Wigner
function can be written as
\begin{eqnarray}
E'(\alpha_1,
\alpha_2,\alpha_3,\alpha_4)&=&\exp\{(-2\cosh2r-\sinh2r) \nonumber
\\
& & \mbox{\hspace{-2cm}}\times
\sum^{4}_{i=1}\alpha_i^2+2\sinh2r\sum^{4}_{i<j}\alpha_i \alpha_j\}
\end{eqnarray}
Evaluate the quantity $\ssb(4)$, and applying the $N$-qubit Bell
inequality, we have
\begin{widetext}
\begin{eqnarray}
\ssb(4)=&&-E'(\alpha_1^1, \alpha_2^1,\alpha_3^1,\alpha_4^1)
+E'(\alpha_1^1, \alpha_2^1,\alpha_3^1,\alpha_4^2)
+E'(\alpha_1^1, \alpha_2^1,\alpha_3^2,\alpha_4^1)+E'(\alpha_1^1, \alpha_2^1,\alpha_3^2,\alpha_4^2)\nonumber \\
&&+E'(\alpha_1^1, \alpha_2^2,\alpha_3^1,\alpha_4^1)
+E'(\alpha_1^1, \alpha_2^2,\alpha_3^1,\alpha_4^2)
+E'(\alpha_1^1, \alpha_2^2,\alpha_3^2,\alpha_4^1)-E'(\alpha_1^1, \alpha_2^2,\alpha_3^2,\alpha_4^2)\nonumber \\
&&+E'(\alpha_1^2, \alpha_2^1,\alpha_3^1,\alpha_4^1)
+E'(\alpha_1^2, \alpha_2^1,\alpha_3^1,\alpha_4^2)+E'(\alpha_1^2, \alpha_2^1,\alpha_3^2,\alpha_4^1)
-E'(\alpha_1^2, \alpha_2^1,\alpha_3^2,\alpha_4^2)\nonumber \\
&&+E'(\alpha_1^2,
\alpha_2^2,\alpha_3^1,\alpha_4^1)-E'(\alpha_1^2,
\alpha_2^2,\alpha_3^1,\alpha_4^2)-E'(\alpha_1^2,
\alpha_2^2,\alpha_3^2,\alpha_4^1)-E'(\alpha_1^2,
\alpha_2^2,\alpha_3^2,\alpha_4^2)
\end{eqnarray}
\end{widetext}
Under a local realistic description, $\ssb(4)\leq4$. By choosing
appropriate measurements, we have $\ssb_{\mbox{opt}}(4)=7.357$.
That is to say that the 4-mode NOPA state shows strong
non-locality, which is stronger compared with 3-mode or 2-mode
NOPA states.
\\
\begin{table}
\begin{tabular}{|c|c|c|c|c|c|c|}
     \hline\hline
$V=2/\ssb_{\mbox{opt}}(N)$
     &   $N=2$       &  $N=3$       &  $N=4$       &  $N=5$  &  $N=6$       &  $N=7$\\
     \hline
ME states
     &   $0.707$  &  $0.5$       &  $0.354$  &  $0.25$ &  $0.177$  &  $0.125$\\
     \hline
Oscillator
     &   $0.913$  &  $0.667$  &  $0.544$  &  $0.4$  &  $0.318$  &  $0.229$\\
     \hline\hline
     \end{tabular}
     \caption{Threshold visibility for $2 \leq N \leq 7$}\label{tab1}\end{table}
\\

\begin{figure}
\begin{center}
\epsfig{figure=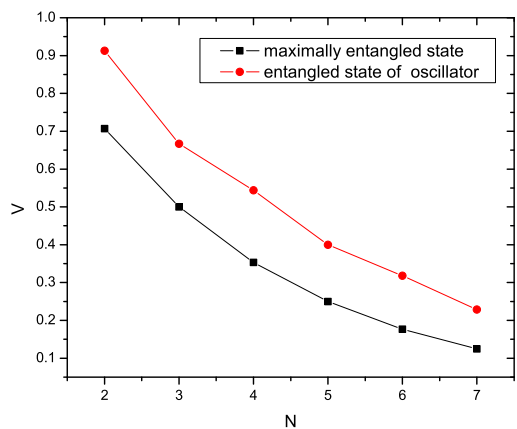,width=0.45\textwidth}
\end{center}
\caption{Critical visibility of N-qubit Bell inequality
(N=2,3,4,5,6,7) for both maximally entangled state and the
entangled state of oscillator.} \label{fig1}
\end{figure}

We also consider the strength of violation or visibility ($V$) as
the minimal amount $V$ of the given entangled state $|\psi\rangle$ that
one has to add to pure noise, $\rho_{\mbox{noise}}$, so that the
resulting state violates local realism. The quantity $V$ is thus the
threshold visibility above which the state cannot be described by
local realism, and it is sometimes called the critical
visibility. More specifically, we consider Werner state of the
form $\rho_w = V |\psi\rangle\langle \psi| +(1-V) \rho_{\mbox{\rm
noise}}$ where $\rho_{\mbox{\rm noise}} = I/2^N$ is the
completely mixed state. As shown in \cite{Zukowski}, for the
maximally entangled state $|\psi\rangle_{\mbox{\rm GHZ}} =
1/\sqrt{2} (|0\rangle_1 \cdots |0\rangle_N + |1\rangle_1 \cdots
|1\rangle_N)$, the Werner state cannot be described by local
realism if and only if $V>1/\sqrt{2^{N-1}}$.

We repeat the calculation for entangled states for $N$
oscillators ($N=2,3,4,5,6,7$) and their results are succinctly
summarized in Table \ref{tab1} and compared to the values for
maximally entangled states. To see the variation of $V$ with $N$,
we also plot $V$ versus the number of particles $N$ both for
maximally entangled (ME) states and entangled states of
oscillators. Naturally it is not surprising to see that the two
systems show similar variations of $V$ with increasing dimension
$N$. Alternatively, if one consider the optimal value of the
violation for the \.{Z}okowski-Brukner inequalities, the optimal
value for this violation grows with $N$. Increasing the number of
qubits, in this case, will not bring us any closer to the
classical regime, but rather it appears to discriminate better
the quantum and the classical boundary. We also see that the
entangled states of the oscillator do not violate the $N$-qubit
Bell inequality as much as the maximally entangled states do
since NOPA state is not maximally entangled. However from the
experimental perspective, NOPA state is easier to generate than
$|\psi\rangle_{\mbox{\rm GHZ}}$.

Our study shows that the multipartite entangled state in the
Wigner representation exhibits non-local realism and this
violation of local realism can be observed using $N$-mode NOPA
state. The violation of local realism for NOPA state can be
manifested through the violation of $N$-particle Bell inequality
for a state described by the Wigner function. This provides an
exciting possibility to test the violation of local realism for
the $N$-mode entangled state experimentally for the general case
using the quantum $N$-mode squeezed state.

We thank Professors B. G. Englert and D. Kaszlikowski for valuable
discussion. This work is supported by NUS academic research grant
WBS: R-144-000-089-112. J.L.C acknowledges financial support from
Singapore Millennium Foundation and (in part) by NSF of China
(No. 10201015).


\begin{thebibliography}{99}
\bibitem{EPR} A. Einstein, B.
Podolsky, and N. Rosen, Phys. Rev. {\bf 47}, 777 (1935).
\bibitem{Bell1} J. S. Bell, Physics, {\bf 1}, 195 (1964).
\bibitem{CH} J.F. Clauser and M.A. Horne, Phys. Rev. D {\bf 10},
526 (1974)
\bibitem{CHSH}  J. F. Clauser, M. A. Horne, A. Shimony
and R. A. Holt, Phys. Rev. Lett. {\bf 23,} 880 (1969).
\bibitem{Green} D. M. Greenberger, M. Horne, A. Shimony, and A.
Zeilinger, Am. J. Phys. {\bf 58}, 1131 (1990).
\bibitem{Peres} A.
Peres, Found. Phys. {\bf 29}, 589 (1999).
\bibitem{JLC} J. L.
Chen, D. Kaszlikowski, L. C. Kwek, and C. H. Oh, Mod. Phys. Lett.
A, {\bf 17}, 2231 (2002); J. L. Chen, D. Kaszlikowski, L. C.
Kwek, C.H. Oh, and M. \.Zukowski, Phys. Rev. A 64, 052109 (2001);
D. Kaszlikowski, L. C. Kwek, J. L. Chen, M. \.Zukowski, and C. H.
Oh Phys. Rev. A 65, 032118 (2002); L.B Fu,J. L Chen, and X. G
Zhao, Phys. Rev. A {\bf 68}, 022323 (2003). L. B. Fu, e-print
quant-ph/0306102; D. Kaszlikowski, D. K. L. Oi, M. Christandl, K.
Chang, A. Ekert, L. C. Kwek, and C. H. Oh, Phys. Rev. A {\bf 67},
012310 (2003); A. Ac\'{i}n, J.L. Chen, N. Gisin, D. Kaszlikowski, L.C. Kwek, C.H. Oh, M. \.{Z}ukowski, Phys. Rev. Lett. {\bf 92}, 250404 (2004); J.L. Chen, C.F. Wu, L.C. Kwek and C.H. Oh, Phys. Rev. Lett. {\bf 93}, 140407 (2004).
\bibitem{Zukowski} M. \.{Z}ukowski and \v{C}.
Brukner, Phys. Rev. Lett. {\bf 88}, 210401 (2002).
\bibitem{Wigner} E. Wigner, Phys. Rev. {\bf 40}, 749 (1932).
\bibitem{Wigner1} M. Hillery, R. F. O'Connell, M. O. Scully, and
E. P. Wigner, Phys. Rep. {\bf 106}, 121 (1984).
\bibitem{Bell2} See, for example, J.S. Bell, {\it Speakable and
unspeakable in quantum mechanics}, (Cambridge University Press,
Camrbidge, 1993).
\bibitem{KK} K. Banaszek and K. W\'odkiewicz, Phys.
Rev. A {\bf 58}, 4345 (1998); Phys. Rev. Lett. {\bf 82,} 2009
(1999); Acta Phys. Slov. {\bf 49 }, 491 (1999).
\bibitem{NOPA} M. D. Reid and P. D. Drummond, Phys. Rev. Lett. {\bf 60,} 2731
(1998).
\bibitem{BGE1} B. G. Englert, J. Phys. A: Math. Gen. {\bf
22}, 625(1989); B. G. Englert, S. A. Fulling, and M. D. Pilloff,
Opt. Commun. 208, 139 (2002).
\bibitem{FAN} H. Y. Fan and N. Q.
Jiang, J. Opt. B: Quantum Semiclass. Opt. {\bf 5}, 283 (2003).
\bibitem{parity} A. Royer, Phys. Rev. A {\bf 15}, 449 (1977); H.
Moya-Cessa and P. L. Knight, Phys. Rev. A {\bf 48}, 2479 (1993).
\end{thebibliography}
\end{document}